\documentclass[12pt,pra,aps,amssymb,amsfonts,amsmath,tightenlines]{revtex4}
\usepackage{dsfont}
\usepackage{graphicx}

\newcommand{\half}{\mbox{$\textstyle{\frac{1}{2}}$}}

\newcommand{\rd}{{\rm d}}
\newcommand{\re}{{\rm e}}

\begin{document}

\title{Optimal Time Evolution for Hermitian and \\ Non-Hermitian
Hamiltonians}

\author{Carl M. Bender${}^*$ and Dorje C. Brody${}^\dagger$}

\affiliation{${}^*$Department of Physics, Washington University, St.
Louis MO 63130, USA\\ ${}^\dagger$Department of Mathematics,
Imperial College London, London SW7 2AZ, UK}

\begin{abstract}

\begin{quotation}
The shortest path between two truths in the real domain passes
through the complex domain. \quad --- Jacques Hadamard, \textit{The
Mathematical
Intelligencer} \textbf{13} (1991)\\
\end{quotation}

Consider the set of all Hamiltonians whose largest and smallest
energy eigenvalues, $E_{\rm max}$ and $E_{\rm min}$, differ by a
fixed energy $\omega$. Given two quantum states, an initial state
$|\psi_I\rangle$ and a final state $|\psi_F\rangle$, there exist
many Hamiltonians $H$ belonging to this set under which
$|\psi_I\rangle$ evolves in time into $|\psi_F\rangle$. Which
Hamiltonian transforms the initial state to the final state in the
least possible time $\tau$? For Hermitian Hamiltonians, $\tau$ has a
nonzero lower bound. However, among complex non-Hermitian
$PT$-symmetric Hamiltonians satisfying the same energy constraint,
$\tau$ can be made arbitrarily small without violating the
time-energy uncertainty principle. The minimum value of $\tau$ can
be made arbitrarily small because for $PT$-symmetric Hamiltonians
the evolution path from the vector $|\psi_I\rangle$ to the vector
$|\psi_F\rangle$, as measured using the Hilbert-space metric
appropriate for this theory, can be made arbitrarily short. The
mechanism described here resembles the effect in general relativity
in which two space-time points can be made arbitrarily close if they
are connected by a wormhole. This result may have applications in
quantum computing.
\end{abstract}
\maketitle

\section{Introduction}
\label{s1}

Interest in optimal time evolution dates back to the end of the
seventeenth century, when the famous brachistochrone problem was
solved almost simultaneously by Newton, Leibniz, l'H\^opital, and
Jacob and Johann Bernoulli. The word \textit{brachistochrone} is
derived from Greek and means shortest time (of flight). The
classical brachistochrone problem is stated as follows: A bead
slides down a frictionless wire from point $A$ to point $B$ in a
homogeneous gravitational field. What is the shape of the wire that
minimises the time of flight of the bead? The solution to this
problem is that the optimal (fastest) time evolution is achieved
when the wire takes the shape of a cycloid, which is the curve that
is traced out by a point on a wheel that is rolling on flat ground.

In the past few years there has been much interest in the
\textit{quantum brachistochrone} problem, which is formulated in a
somewhat similar fashion: Consider two fixed quantum states, an
initial state $|\psi_I\rangle$ and a final state $|\psi_F\rangle$ in
a Hilbert space. We then consider the set of all Hamiltonians
satisfying the energy constraint that the difference between the
largest and smallest eigenvalues is a fixed energy $\omega$: $E_{\rm
max}-E_{\rm min}=\omega$. Some of the Hamiltonians in this set allow
the initial state $|\psi_I \rangle$ to evolve into the final state
$|\psi_F\rangle$ in time $t$:
\begin{equation}
|\psi_F\rangle=\re^{-{\rm i}Ht/\hslash}|\psi_I\rangle.
\label{e1}
\end{equation}
The quantum brachistochrone problem is to find the \textit{optimal}
Hamiltonian; that is, the Hamiltonian that accomplishes this
evolution in the shortest possible time, which we denote by $\tau$.

In this chapter we show that for Hermitian Hamiltonians the shortest
evolution time $\tau$ is a nonzero quantity whose size depends on
the Hilbert-space distance between the fixed initial and final state
vectors. However, for complex non-Hermitian Hamiltonians, the value
of $\tau$ can be made arbitrarily small. Thus, non-Hermitian
Hamiltonians permit arbitrarily fast time evolution.

Of course, a non-Hermitian Hamiltonian may be physically unrealistic
because it may possess complex eigenvalues and/or it may generate
nonunitary time evolution; that is, time evolution in which
probability is not conserved. However, there is a special class of
non-Hermitian Hamiltonians that are $PT$ symmetric; that is,
Hamiltonians that are invariant under combined space and time
reflection. Although such Hamiltonians are not Hermitian in the
Dirac sense, they \textit{do} have entirely real spectra and give
rise to unitary time evolution. Thus, such Hamiltonians define
consistent and acceptable theories of quantum mechanics. We show in
this chapter that if we use Hamiltonians of this type to solve the
quantum brachistochrone problem, we can achieve arbitrarily fast
time evolution without violating any principles of quantum
mechanics. Thus, if it were possible to implement
faster-than-Hermitian time evolution, then non-Hermitian
Hamiltonians might have important applications in quantum computing.

This chapter is organised as follows: In Sec.~\ref{s2} we introduce
and describe $PT$ quantum mechanics and explain how a Hamiltonian
that is not Dirac Hermitian can still define a consistent theory of
quantum mechanics. Then in Sec.~\ref{s3} we explain why complex
classical mechanics allows for faster-than-conventional time
evolution. In Sec.~\ref{s4} we discuss the quantum brachistochrone
for Hermitian Hamiltonians. Then, in Sec.~\ref{s5} we extend the
discussion in Sec.~\ref{s4} to Hamiltonians that are not Dirac
Hermitian. In Sec.~\ref{s6} we explain how it might be possible for
a complex Hamiltonian to achieve faster-than-Hermitian time
evolution.

\section{$PT$ Quantum Mechanics}
\label{s2}

Based on the training that one receives in a traditional quantum
mechanics course, one would expect a theory defined by a
non-Hermitian Hamiltonian to be physically unacceptable because the
energy levels would most likely be complex and the time evolution
would most likely be nonunitary (not probability-conserving).
However, theories defined by a special class of non-Hermitian
Hamiltonians called $PT$-symmetric Hamiltonians can have positive
real energy levels and can exhibit unitary time evolution. Such
theories are consistent quantum theories. It may be possible to
distinguish these theories experimentally from theories defined by
Hermitian Hamiltonians because, in principle, non-Hermitian
Hamiltonians can be used to generate arbitrarily fast time
evolution.

We use the following terminology in this chapter: By
\textit{Hermitian}, we mean \textit{Dirac} Hermitian, where the
Dirac Hermitian adjoint symbol $^\dag$ represents combined matrix
transposition and complex conjugation. The \textit{parity operator}
$P$ performs spatial reflection and thus in quantum mechanics it
changes the sign of the position operator $x$ and the momentum
operator $p$: $PxP=-x$ and $PpP=-p$. Because the parity operator $P$
is a reflection operator, its square is the unit operator:
$P^2={\mathds 1}$. The \textit{time-reversal operator} $T$ performs
the time reflection $t\to-t$, and thus it changes the sign of the
momentum operator $p$, $TpT=-p$, but it leaves the position operator
invariant: $TxT=x$. The square of $T$ is the unit operator
$T^2={\mathds 1}$. We require that the operators $P$ and $T$
individually leave invariant the fundamental Heisenberg algebra of
quantum mechanics $[x,p]={\rm i}$. Thus, while $P$ is a linear
operator, we see that $T$ must perform \textit{complex conjugation}
$TzT=z^*$, and hence $T$ is an \textit{antilinear}
operator.\footnote{Another way to see that $T$ is associated with
complex conjugation is to require that the time-dependent
Schr\"odinger equation be invariant under time reversal. This
implies that time reflection $t\to-t$ must be accompanied by complex
conjugation ${\rm i}\to-{\rm i}$. See Ref.~\cite{X7} for a
discussion of the properties of antilinear operators.}

The first class of $PT$-symmetric quantum-mechanical Hamiltonians
was introduced in 1998 \cite{R1}. Since then there have been many
papers on this subject by a wide range of authors. There have also
been three recent review articles \cite{R2,R3,R4}. In Ref.~\cite{R1}
it was discovered that even if a Hamiltonian is not Hermitian, its
energy levels can be all real and positive so long as the
eigenfunctions are symmetric under $PT$ reflection.

These new kinds of Hamiltonians are obtained by deforming ordinary
Hermitian Hamiltonians into the complex domain. The original class
of $PT$-symmetric Hamiltonians that was proposed in Ref.~\cite{R1}
has the form
\begin{equation}
H=p^2+x^2({\rm i}x)^\epsilon\qquad(\epsilon>0),
\label{e2}
\end{equation}
where $\epsilon$ is a real deformation parameter. Two particularly
interesting special cases are obtained by setting $\epsilon=1$ to
get $H=p^2+{\rm i}x^3$ and by setting $\epsilon=2$ to get
$H=p^2-x^4$. Surprisingly, these Hamiltonians have real, positive,
discrete energy levels even though the potential for $\epsilon=1$ is
imaginary and the potential for $\epsilon=2$ is upside-down. The
first complete proof of spectral reality and positivity for $H$ in
(\ref{e2}) was given by Dorey \textit{et al.} in Ref.~\cite{R5,R6}.

The philosophical background of $PT$ quantum mechanics is simply
this: One of the axioms of quantum mechanics requires that the
Hamiltonian $H$ be Dirac Hermitian. This axiom is distinct from all
other quantum-mechanical axioms because it is mathematical rather
than physical in character. The other axioms of quantum mechanics
are stated in physical terms; these other axioms require locality,
causality, stability and uniqueness of the vacuum state,
conservation of probability, Lorentz invariance, and so on. The
condition of Dirac Hermiticity $H=H^\dag$ is mathematical, but the
condition of $PT$ symmetry $H=H^{PT}=(PT)H(PT)$ (space-time
reflection symmetry) is physical because $P$ and $T$ are elements of
the Lorentz group.

The spectrum of $H$ in (\ref{e2}) is real, which poses the question
of whether this Hamiltonian specifies a \textit{quantum-mechanical}
theory. That is, is the theory specified by $H$ associated with a
Hilbert space endowed with a positive inner product and does $H$
specify unitary (norm-preserving) time evolution? The answer to
these questions is \textit{yes}. Positivity of the inner product and
unitary time evolution was established in Ref.~\cite{R7,R8} for
quantum-mechanical systems having an unbroken $PT$ symmetry (an
analogous result was obtained by Mostafazadeh in Ref.~\cite{X9}) and
in Ref.~\cite{R9} for quantum field theory.

To demonstrate that the theory specified by the $H$ in (\ref{e2}) is
a quantum-mechanical theory, we construct a linear operator $C$ that
satisfies the three simultaneous algebraic equations \cite{R7}:
$C^2={\mathds 1}$, $[C,PT]=0$, and $[C,H]=0$. Using $C$, which in
quantum field theory is a Lorentz scalar \cite{R10}, we can then
construct the appropriate inner product for a $P T$-symmetric
Hamiltonian: $\langle a|b\rangle\equiv a^{CPT}\cdot b$. This inner
product, which uses the $CPT$ adjoint, has a strictly positive norm:
$\langle a| a\rangle>0$. Because $H$ commutes with both $PT$ and
$C$, $H$ is \textit{self-adjoint} with respect to $CPT$ conjugation.
Also, the time-evolution operator $\re^{-{\rm i}Ht/\hslash}$ is
unitary with respect to $C PT$ conjugation. Note that the Hilbert
space and the $CPT$ inner product are \textit{dynamically
determined} by the Hamiltonian itself.

We have explained why a $PT$-symmetric Hamiltonian gives rise to a
unitary theory, but in doing so we raise the question of whether
$PT$-symmetric Hamiltonians are useful. The answer to this question
is simply that $P T$-symmetric Hamiltonians have \textit{already}
been useful in many areas of physics. For example, in 1959 Wu showed
that the ground state of a Bose system of hard spheres is described
by a non-Hermitian Hamiltonian \cite{R11}. Wu found that the
ground-state energy of this system is real and he conjectured that
all of the energy levels were real. Hollowood showed that the
non-Hermitian Hamiltonian for a complex Toda lattice has real energy
levels \cite{R12}. Cubic non-Hermitian Hamiltonians of the form
$H=p^2+{\rm i}x^3$ (and also cubic quantum field theories having an
imaginary self-coupling term) arise in studies of the Lee-Yang edge
singularity \cite{R14,R15,R13,R16} and in various Reggeon
field-theory models \cite{R17,R18,R19}. In all of these cases a
non-Hermitian Hamiltonian having a real spectrum appeared mysterious
at the time, but now the explanation is clear: In every case the
non-Hermitian Hamiltonian is $PT$ symmetric. Hamiltonians having
$PT$ symmetry have also been used to describe magnetohydrodynamic
systems \cite{R21,R20} and to study nondissipative time-dependent
systems interacting with electromagnetic fields \cite{R22}.

An important application of $PT$ quantum mechanics is in the
revitalisation of theories that have been thought to be dead because
they appear to have ghosts. \textit{Ghosts} are states having
negative norm. We have explained above that in order to construct
the quantum-mechanical theory defined by a $PT$-symmetric
Hamiltonian, we must construct the appropriate adjoint from the $C$
operator. Having constructed the $CPT$ adjoint, one may find that
the so-called ghost state is actually not a ghost at all because
when its norm is calculated using the appropriate definition of the
adjoint, the norm turns out to be positive. This is what happens in
the case of the Lee model.

The Lee model was proposed in 1954 as a quantum field theory in
which mass, wave-function, and charge reorganisation could be
performed exactly and in closed form \cite{R23}. However, in 1955
K\"all\'en and Pauli showed that when the renormalised coupling
constant is larger than a critical value, the Hamiltonian becomes
non-Hermitian (in the Dirac sense) and a ghost state appears
\cite{R24}. The importance of the work of K\"all\'en and Pauli was
emphasised by Salam in his review of their paper \cite{R58} and the
appearance of the ghost was assumed to be a fundamental defect of
the Lee model. However, in 2005 it was shown that the non-Hermitian
Lee-model Hamiltonian is $PT$ symmetric and when the norms of the
states of this model are determined using the $C$ operator, which
can be calculated exactly and in closed form, the ghost state is
seen to be an ordinary physical state having positive norm
\cite{R25}. Thus, the following assertion by Barton \cite{R26} is
\textit{not correct}: ``A non-Hermitian Hamiltonian is unacceptable
partly because it may lead to complex energy eigenvalues, but
chiefly because it implies a non-unitary S matrix, which fails to
conserve probability and makes a hash of the physical
interpretation.''

Another example of a quantum model that was thought to have ghost
states, but in fact does not, is the Pais-Uhlenbeck oscillator model
\cite{R27,R28,R29}. This model has a fourth-order field equation,
and for the past several decades it was thought (incorrectly) that
all such higher-order field equations lead inevitably to ghosts. It
is shown in Ref.~\cite{R27} that when the Pais-Uhlenbeck model is
quantised using the methods of $PT$ quantum mechanics, it does not
have any ghost states at all.

There are many potential applications for $PT$ quantum mechanics in
areas such as particle physics, cosmology, gravitation, quantum
field theory, and solid-state physics. These applications are
discussed in detail in the recent review article in Ref.~\cite{R3}.
Furthermore, there are now indications that theories described by
$PT$-symmetric Hamiltonians can be observed in table-top experiments
\cite{R31,R30,R32}.

Having shown the validity and potential usefulness of $PT$ quantum
mechanics, one may ask why $PT$ quantum mechanics works. The reason
is that $CP$ is a positive operator, and thus it can be written as
the exponential of another operator $Q$: $CP=\re^Q$. The square root
of $\re^Q$ can then be used to construct a new Hamiltonian $\tilde
H$ via a similarity transformation on the $P T$-symmetric
Hamiltonian $H$: $\tilde H\equiv\re^{-Q/2}H\re^{Q/2}$. The new
Hamiltonian $\tilde H$ has the same energy eigenvalues as the
original Hamiltonian $H$ because a similarity transformation is
isospectral. Moreover, $\tilde H$ is \textit{Dirac Hermitian}
\cite{R33}; $PT$ quantum mechanics works because there is an
isospectral equivalence between a non-Hermitian $P T$-symmetric
Hamiltonian and a conventional Dirac Hermitian Hamiltonian.

There are a number of elementary examples of this equivalence, but a
nontrivial illustration is provided by the Hamiltonian $H$ in
(\ref{e2}) at $\epsilon=2$. This Hamiltonian is not Hermitian
because boundary conditions that violate the $L^2$ norm must be
imposed in Stokes wedges in the complex plane in order to obtain a
real, positive, discrete spectrum. The exact equivalent Hermitian
Hamiltonian is $\tilde H=p^2+4x^4-2\hslash\,x$, where $\hslash$ is
Planck's constant \cite{R36,R34,R35}. The term proportional to
$\hslash$ vanishes in the classical limit and is thus an example of
a quantum anomaly.

Since $PT$ symmetry is equivalent by means of a similarity
transformation to conventional Dirac Hermiticity, one may wonder
whether $PT$ quantum mechanics is actually fundamentally different
from ordinary quantum mechanics. The answer is \textit{yes} and, at
least in principle, there is an experimentally observable difference
between $PT$-symmetric and ordinary Dirac Hermitian Hamiltonians.
The quantum brachistochrone provides a setting for examining this
difference and provides a way to discriminate between the class of
$PT$-symmetric Hamiltonians and the class of Dirac Hermitian
Hamiltonians.

\section{Complex Classical Motion}
\label{s3}

It is implicitly assumed in the derivation of the classical
brachistochrone that the path of shortest time of descent is
\textit{real}. However, it is interesting that if one allows for the
possibility of complex paths of motion, one can achieve an even
shorter time of flight. In this section we consider a simple
classical-mechanical system. Our purpose is to explain heuristically
how extending a dynamical system into the complex domain can result
in faster-than-real time evolution.

\begin{figure}[!th]
\begin{center}
  \includegraphics[scale=0.6,angle=-90]{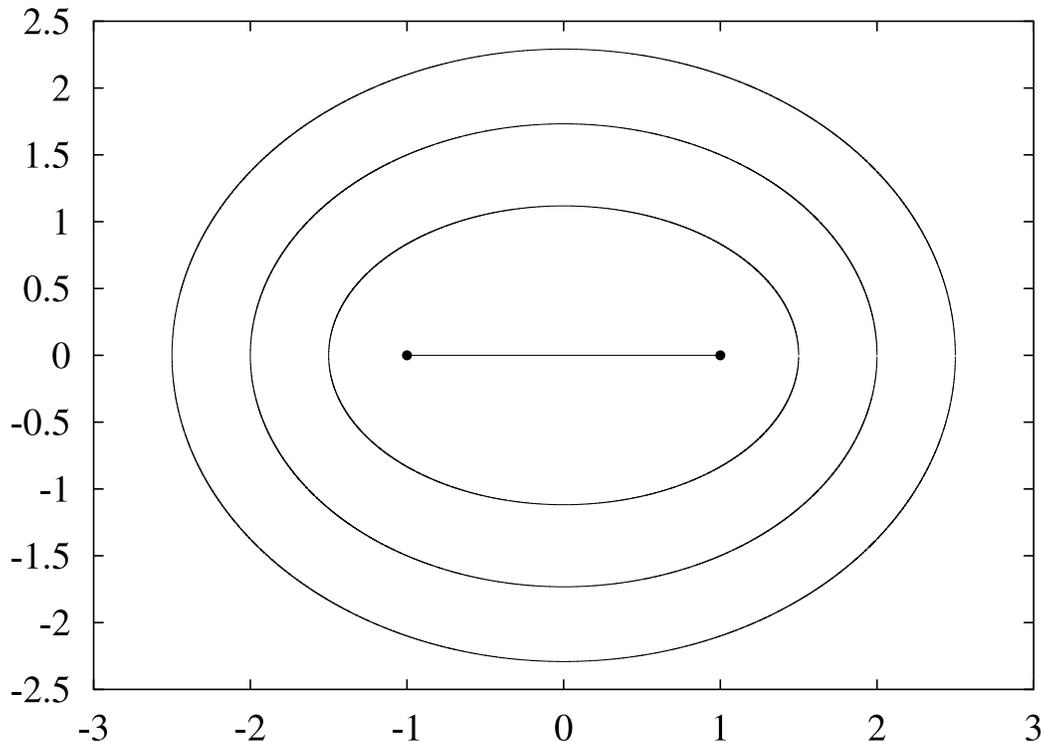}
  \caption{Classical trajectories in
the complex-$x$ plane for the harmonic oscillator whose Hamiltonian
is $H=p^2+x^2$. These trajectories represent the possible paths of a
particle whose energy is $E=1$. The trajectories are nested ellipses
with foci located at the turning points at $x=\pm1$. The real line
segment (degenerate ellipse) connecting the turning points is the
usual periodic classical solution to the harmonic oscillator. All
closed paths have the same period $2\pi$.
  \label{f1}
  }
\end{center}
\end{figure}

To demonstrate that a shorter time of flight can be achieved by
means of complex paths, let us consider the classical harmonic
oscillator, whose Hamiltonian is
\begin{equation}
H=p^2+x^2.
\label{e3}
\end{equation}
If a particle has energy $E=1$, then the classical turning points of
the motion of the particle are located at $x=\pm1$. The particle
undergoes simple harmonic motion in which it oscillates sinusoidally
between these two turning points. This periodic motion is indicated
in Fig.~\ref{f1} by a solid line connecting the turning points.
However, in addition to this oscillatory motion on the real-$x$
axis, there are an infinite number of other trajectories that a
particle of energy $E$ can have \cite{R37}. These classical
trajectories, which are also shown in Fig.~\ref{f1}, are all
ellipses whose foci are located at precisely the positions of the
turning points. All of the classical orbits are periodic, and all
orbits have the same period $T=2\pi$. Thus, a classical particle
travels faster along more distant ellipses.

Now suppose that a classical particle of energy $E=1$ travels along
the real-$x$ axis from some point $x=-a$ to $x=a$, where $a>1$. If
the potential $V(x)$ is everywhere zero along its path, then it will
travel at a constant velocity. However, suppose that the particle
suddenly finds itself in the parabolic potential $V(x)=x^2$ when it
reaches the turning point at $x=-1$ and that it suddenly escapes the
influence of this potential at $x=1$. Then, the time of flight from
$x=-a$ to $x=a$ will be changed because the particle does not travel
at a constant velocity between the turning points. Next, let us
imagine that the potential $V(x)=x^2$ is suddenly turned on
\textit{before} the particle reaches the turning point at $x=-1$. In
this case, the particle will follow one of the elliptical paths in
the complex plane around to the positive real axis. Just as the
particle reaches the positive real axis the potential is turned off,
so the particle proceeds onward along the real axis until it reaches
$x=a$. This trip will take less time because the particle travels
faster along the ellipse in the complex plane.

We have arrived at the surprising conclusion that if the classical
particle enters the parabolic potential $V(x)=x^2$ immediately after
it begins its voyage up the real axis, its time of flight will be
exactly half a period, or $\pi$. Indeed, by travelling in the
complex plane, a particle of energy $E=1$ can go from the point
$x=-a$ to the point $x=a$ in time $\pi$, no matter how large $a$ is.
Evidently, if a particle is allowed to follow complex classical
trajectories, then it is possible to make a drastic reduction in its
time of flight between two given real points.

\section{Hermitian Quantum Brachistochrone}
\label{s4}

The quantum brachistochrone problem, as described briefly in
Sec.~\ref{s1}, is similar to the classical counterpart except that
the optimisation takes place in a Hilbert space. Specifically, we
are given a pair of quantum states, an initial state
$|\psi_I\rangle$ and the final state $|\psi_F\rangle$, and we would
like to find the one-parameter family of unitary operators $\{U_t\}$
that achieves the transformation
$|\psi_I\rangle\to|\psi_F\rangle=U_t|\psi_I\rangle$ in the smallest
possible time $t$. Since a one-parameter family of unitary operators
can be formed in terms of a Hermitian operator $H$ as
$U_t=\exp(-{\rm i}Ht/\hslash)$, the problem is equivalent to finding
the Hermitian operator $H$ that realises the transformation
$|\psi_I\rangle\to|\psi_F\rangle$ in the shortest possible time.

The Hermitian operator $H$ can be thought of as representing the
Hamiltonian, so the quantum brachistochrone problem is equivalent to
finding the optimal Hamiltonian $H$ satisfying $\exp(-{\rm
i}Ht/\hslash)|\psi_I\rangle=|\psi_F \rangle$. However, it is
intuitively clear that if we are allowed to have access to an
unbounded energy resource, then the time required for the relevant
transformation, irrespective of whether the Hamiltonian is optimal
or not, can be made arbitrary small. Hence, for a quantum
brachistochrone problem to possess a nontrivial solution, some form
of constraint is needed. The simplest constraint is to assume that
the energy is bounded so that the difference between the largest and
the smallest energy eigenvalues has a fixed value $\omega$:
\begin{eqnarray}
E_{\rm max}-E_{\rm min}=\omega.
\label{e4}
\end{eqnarray}
A short calculation shows that if the Hamiltonian $H$ is bounded,
then (i) the standard deviation of the energy is bounded according
to
\begin{eqnarray}
\Delta H\leq\half(E_{\rm max}-E_{\rm min}),
\label{e5}
\end{eqnarray}
and (ii) the state with maximum energy uncertainty is $(|E_{\rm
max}\rangle+| E_{\rm min}\rangle)/\sqrt{2}$. It follows that the
energy constraint (\ref{e4}) is equivalent to a constraint on energy
uncertainty.

The brachistochrone problem of this type has been analysed recently
and a solution was obtained by means of a variational method
\cite{R39}. It has also been solved in terms of a more elementary
approach making use of the geometry of quantum state space
\cite{R40}. We shall follow closely the latter approach here.

Let us now state the simplest form of the quantum brachistochrone
problem: We have a quantum system represented by an $N$-dimensional
Hilbert space ${\mathcal H}$ and a prescribed pair of states
$|\psi_I\rangle$ and $|\psi_F\rangle$ on ${\mathcal H}$. The problem
is (a) to find the Hamiltonian $H$ satisfying the constraint
(\ref{e4}) such that the unitary transformation $\exp(-{\rm i}Ht/
\hslash)|\psi_I\rangle=|\psi_F\rangle$ is achieved in shortest
possible time and (b) to find the time required to realise such an
operation.

A little geometric intuition allows us to find the solution to this
problem with minimum effort. Recall that the time required for
transporting a state along a path in ${\mathcal H}$ is given by the
\textit{distance} divided by the \textit{speed}. Hence, all we have
to do is first to identify the shortest path and measure its length
and then to allow the state to evolve along the path with the
greatest possible speed without violating the constraint (\ref{e4}).

In quantum mechanics the notion of distance is closely linked to the
notion of transition probability \cite{R43,R44}. In particular, by
looking at the transition probability between neighbouring states we
can derive the expression for the metric on the space of quantum
states. This allows us to measure distances between states. The idea
can be sketched as follows: Consider a state $|\psi\rangle$ in
${\mathcal H}$ and a neighbouring state $|\psi\rangle+|\rd\psi
\rangle$. The transition probability between these states is
\begin{eqnarray}
\cos^2\left(\half\rd
s\right)=\frac{(\langle\psi|+\langle\rd\psi|)|\psi\rangle
\langle\psi|(|\psi\rangle+|\rd\psi\rangle)}{\langle\psi
|\psi\rangle(\langle\psi| +\langle
\rd\psi|)(|\psi\rangle+|\rd\psi\rangle)}, \label{e6}
\end{eqnarray}
where $\rd s$ defines the line element on the space of pure states.
By using
\begin{equation}
\cos^2\left(\half\rd s\right)\approx1-\frac{1}{4}\rd s^2,
\label{e6.5}
\end{equation}
expanding the right side of (\ref{e6}), and retaining terms of
quadratic order, we find that the line element is
\begin{eqnarray}
\rd s^2=4\frac{\langle\psi|\psi\rangle\langle\rd\psi|\rd\psi\rangle
-\langle\psi| \rd\psi\rangle\langle\rd\psi|\psi\rangle}
{\langle\psi|\psi\rangle^2}. \label{e7}
\end{eqnarray}
This line element is known in geometry to arise from the
\textit{Fubini-Study metric} \cite{R48} and it can be used to
measure the distance of the shortest path joining a pair of points
on the space of pure quantum states.

If the Hilbert space is two dimensional, then a generic normalised
state vector $|\psi\rangle$ can be expressed in the form
\begin{eqnarray}
|\psi\rangle=\left(\begin{array}{c}\cos\frac{1}{2}\theta\\ \\
\sin\frac{1}{2}\theta\,\re^{{\rm i}\phi}\end{array}\right).
\label{e8}
\end{eqnarray}
A short calculation then shows that the Fubini-Study line element
(\ref{e7}) reduces in this case to the expression
\begin{eqnarray}
\rd s^2=\rd\theta^2+\sin^2\theta\,\rd\phi^2,
\label{e9}
\end{eqnarray}
which we recognise as the line element on the Bloch sphere
${\mathcal S}$. (The Bloch sphere is the state space of two-level
systems.)

In the case of an $n$-dimensional Hilbert space ${\mathcal H}$, if
we are given a pair of distinct states $|\psi_I\rangle$ and
$|\psi_F\rangle$, then the linear span of these two states forms a
two-dimensional subspace of ${\mathcal H}$. It should be intuitively
clear that the shortest path joining $|\psi_I\rangle$ and
$|\psi_F\rangle$ should lie on this two-dimensional subspace. Thus,
irrespective of the dimensionality of ${\mathcal H}$ \textit{the
solution to our quantum brachistochrone problem can be obtained by
analysing the two-dimensional subspace spanned by $|\psi_I\rangle$
and $|\psi_F\rangle$}. Even when we restrict our attention to this
subspace, there still remain infinitely many unitary orbits that
realise the transformation $|\psi_I\rangle\to|\psi_F\rangle=
U_t|\psi_I\rangle$. However, since the two-dimensional state space
is just the Bloch sphere ${\mathcal S}$ endowed with the spherical
metric (\ref{e9}), we see that there is a unique great circle arc
that joins $|\psi_I\rangle$ and $|\psi_F \rangle$ on ${\mathcal S}$.
(This assumes, of course, that $|\psi_I\rangle$ and $|\psi_F\rangle$
are not antipodal points of the sphere. Otherwise, there are
infinitely many such paths.) In this way we have identified the
shortest path joining $|\psi_I\rangle$ and $|\psi_F\rangle$. The
shortest distance $s_{\rm min}$ between these two points of
${\mathcal S}$ is thus given by
\begin{eqnarray}
s_{\rm min}=2\arccos\left(\frac{|\langle\psi_I|\psi_F\rangle|}
{\sqrt{\langle \psi_I|\psi_I\rangle\langle\psi_F|\psi_F\rangle}}
\right). \label{e10}
\end{eqnarray}
This result can also be obtained by integrating the line element
(\ref{e9}) along the great-circle arc on ${\cal S}$.

Having obtained the distance of the shortest path we proceed to find
the maximum speed at which the state can evolve unitarily. For the
evolution of the state we must consider the general Schr\"odinger
equation, but we also need to express the equation in the correct
form. This is the so-called modified Schr\"odinger equation
\begin{eqnarray}
\frac{\rd|\psi_t\rangle}{\rd t}=-\frac{\rm i}{\hslash}\,{\tilde H}
|\psi_t \rangle. \label{e11}
\end{eqnarray}
In this equation the mean-adjusted Hamiltonian ${\tilde H}$ is given
by
\begin{eqnarray}
{\tilde H}=H-\langle H\rangle,
\label{e12}
\end{eqnarray}
where
\begin{eqnarray}
\langle
H\rangle=\frac{\langle\psi|H|\psi\rangle}{\langle\psi|\psi\rangle}.
\label{e12.1}
\end{eqnarray}
Note that $\langle{\tilde H}\rangle=0$ and that according to
(\ref{e11}) the tangent vector $\frac{\rd}{\rd t}|psi_t\rangle$ is
everywhere orthogonal to the direction of the state \cite{K2}. Since
the energy expectation $\langle H \rangle$ depends on the state
$|\psi\rangle$, the modified Schr\"odinger equation appears to be
nonlinear. However, this is not the case. The point is that the
expectation value of the Hamiltonian is a constant of the motion
under the Schr\"odinger dynamics. Thus, given the initial state
$|\psi_I\rangle$, we calculate $\langle H\rangle$ in this state and
subtract this number from the Hamiltonian. Since the Hamiltonian in
quantum mechanics is defined only up to an additive constant, this
modification does not alter the physics in any way. It is worthwhile
noting that the modified Schr\"odinger equation (\ref{e11}) is
canonical and reduces to the standard eigenvalue problem when the
state $|\psi_t \rangle$ is time independent without one having to
evoke the correspondence principle \cite{R43}.

If the initial state vector $|\psi_I\rangle$ is normalised, then the
evolution (\ref{e11}) preserves the norm. It follows that
$\langle\psi|\rd\psi\rangle=0$. Since the speed $v$ of quantum
evolution is given by $v=\rd s/\rd t$, we find from (\ref{e7}) and
(\ref{e11}) that
\begin{eqnarray}
v^2=\frac{4}{\hslash^2}\,\langle\psi_t|(H-\langle
H\rangle)^2|\psi_t\rangle=
\frac{4}{\hslash^2}\,\langle\psi_I|(H-\langle
H\rangle)^2|\psi_I\rangle. \label{e13}
\end{eqnarray}
This shows that the speed of quantum evolution is given by the
energy uncertainty. The expression (\ref{e13}) for the speed of
quantum evolution is known as the \textit{Anandan-Aharonov relation}
\cite{R47}. Since we know from (\ref{e5}) that under the constraint
(\ref{e4}) the energy uncertainty $\Delta H$ is bounded by $\half
\omega$, we find that the maximum speed of quantum evolution is
given by
\begin{eqnarray}
v_{\rm max}=\frac{\omega}{\hslash}.
\label{e14}
\end{eqnarray}
By using the results in (\ref{e10}) and (\ref{e14}) we deduce that
the minimum time required for realising the unitary transportation
$|\psi_I\rangle\to|\psi_F \rangle=U_t|\psi_I\rangle$ is given by the
ratio $s_{\rm min}/v_{\rm max}$. In particular, if $|\psi_I\rangle$
and $|\psi_F\rangle$ are orthogonal, then they correspond to
antipodal points on the Bloch sphere ${\mathcal S}$, and we have
$s_{\rm min}=\pi$. In this case, the minimum time required to
orthogonalize the state (that is, for the state to evolve into a new
state that is orthogonal to the original state) is known as the
\textit{passage time} $\tau_{\rm P}$ \cite{R42,R46}. The passage
time is explicitly
\begin{eqnarray}
\tau_{\rm P}=\frac{\pi\hslash}{2\Delta H}=\frac{\pi\hslash}{\omega}.
\label{e15}
\end{eqnarray}
The passage time (\ref{e15}) provides the bound in Hermitian quantum
mechanics for transporting a state into an orthogonal state, and is
sometimes referred to as the \textit{Fleming bound} \cite{R45}.

\begin{figure}[t]
\begin{center}
\includegraphics[scale=.75]{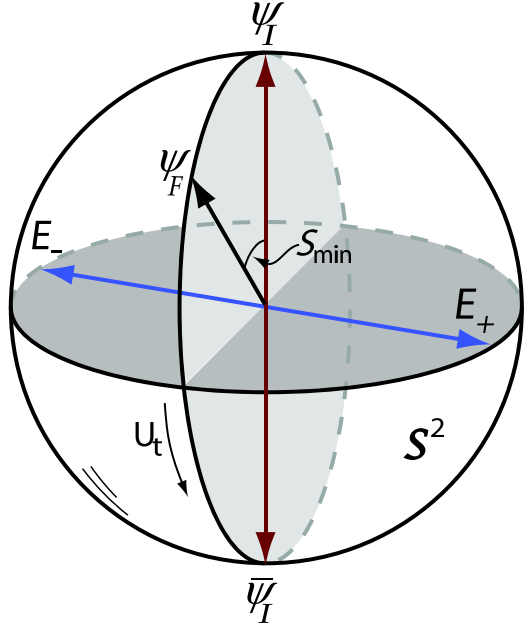}
\end{center}
\caption{Optimal Hamiltonian for quantum state transportation. The
two-dimensional complex Hilbert space spanned by the initial state
$|\psi_I \rangle$ and the final state $|\psi_F\rangle$ can be
visualised in real terms as a Bloch sphere ${\mathcal S}$. The two
states $|\psi_I\rangle$ and $|\psi_F \rangle$ can then be identified
as a pair of points on ${\mathcal S}$. Assuming that these two
points are not antipodal, there exists a unique great circle arc
joining these two points, which determines the shortest path joining
the two states. The optimal way of unitarily transporting
$|\psi_I\rangle$ into $|\psi_F \rangle$ is therefore to rotate the
sphere along the axis orthogonal to the great circle. The axis of
rotation then specifies two quantum states: $|E_- \rangle$ and
$|E_+\rangle$. These states are the eigenstates of the optimal
Hamiltonian $H$.} \label{fig:1}
\end{figure}

The ratio $s_{\rm min}/v_{\rm max}$ gives the solution to part (b)
of our quantum brachistochrone problem. To solve part (a), that is,
to find the optimal Hamiltonian, we argue as follows: Since the
problem is confined to a two-dimensional subspace of ${\mathcal H}$,
the solution can be obtained by elementary trigonometry on the Bloch
sphere ${\mathcal S}$. The key idea is to recall that the shortest
path joining $|\psi_I\rangle$ and $|\psi_F\rangle$ is a great circle
arc on ${\mathcal S}$. Without loss of generality, we can assume
that $|\psi_I\rangle$ and $|\psi_F\rangle$ lie on the equator of
${\mathcal S}$ with respect to a suitable choice of axis. Then, the
unitary motion along the shortest path can be generated by the
rotation of ${\mathcal S}$ along this axis. Since the eigenstates of
the Hamiltonian $H$ that generates such a rotation correspond to the
antipodal points along this axis, the pair of states
$|\psi_I\rangle$ and $|\psi_F\rangle$ can both be expressed as equal
superpositions of the eigenstates of $H$ with the relative phase
shifted by $s_{\rm min}$. Writing $|E_+\rangle$ and $|E_-\rangle$
for the normalised eigenstates of $H$ and using $\alpha=s_{\rm
min}/2$, we can express the initial and the final state in the form
\begin{eqnarray}
{\textstyle\frac{1}{\surd2}}\big(|E_-\rangle+\re^{-{\rm
i}\alpha}|E_+\rangle \big)=|\psi_I\rangle\quad{\rm
and}\quad{\textstyle\frac{1}{\surd2}}\big(|E_- \rangle+ \re^{{\rm
i}\alpha}|E_+\rangle\big)=|\psi_F\rangle. \label{e16}
\end{eqnarray}
Solving these equations for $|E_+ \rangle$ and $|E_-\rangle$, we obtain
\begin{eqnarray}
|E_-\rangle=\frac{{\rm i}}{{\surd2}\sin\alpha}\big(\re^{-{\rm
i}\alpha}|\psi_F \rangle-\re^{{\rm i}\alpha}|\psi_I\rangle\big)
\label{e17}
\end{eqnarray}
and
\begin{eqnarray}
|E_+\rangle=-\frac{{\rm
i}}{{\surd2}\sin\alpha}\big(|\psi_F\rangle-|\psi_I \rangle\big).
\label{e18}
\end{eqnarray}
These are the eigenstates of the optimal Hamiltonian $H$ that
generates the unitary motion
$|\psi_I\rangle\to|\psi_F\rangle=U_t|\psi_I\rangle$ along the
shortest path. The eigenvalues of the optimal $H$ can be arbitrary
as long as the condition (\ref{e4}) is satisfied. Without loss of
generality, we may assume $H$ be trace free, and we obtain the
solution to part (a) of the quantum brachistochrone problem:
\begin{eqnarray}
H=\half\omega|E_+\rangle\langle E_+|-\half\omega|E_-\rangle\langle
E_-|. \label{e19}
\end{eqnarray}
This is the `minimal' solution to the problem in the sense that $H$
acts only on the two-dimensional subspace of ${\mathcal H}$ while
leaving the rest of ${\mathcal H}$ unperturbed.

As a special example, consider the problem of a spin-flip, that is,
turning a spin-up state into a spin-down state unitarily. In this
case the initial and the final states can be written as
\begin{eqnarray}
|\psi_I\rangle=\left({1\atop0}\right)\quad{\rm
and}\quad|\psi_F\rangle=\left( {0\atop1}\right) \label{e20}
\end{eqnarray}
in the spin-$z$ basis. Substituting these into (\ref{e17}) and
(\ref{e18}), we find that the eigenstates of the optimal Hamiltonian
are
\begin{eqnarray}
|E_-\rangle=\frac{1}{\surd2}\left({1\atop1}\right)\quad{\rm
and}\quad|E_+\rangle =\frac{{\rm i}}{\surd2}\left({1\atop-1}\right)
\label{e21}
\end{eqnarray}
because in this case we have $\alpha=\pi/2$. Substituting this result into
(\ref{e19}) yields
\begin{eqnarray}
H=\half\left(\begin{array}{cc}0&-\omega\\ -\omega & 0\end{array}\right)
\label{e22}
\end{eqnarray}
for the optimal Hamiltonian. By using the relation
\begin{eqnarray}
\re^{{\rm i}\phi{\bf\sigma}\cdot{\bf n}} = \cos\phi
{\mathds 1}+{\rm i}\sin\phi {\bf\sigma}\cdot{\bf n},
\label{e23}
\end{eqnarray}
where ${\bf n}$ is a unit vector and
\begin{equation}
\sigma_1=\left(\begin{array}{cc}0&1\\ 1&0\end{array}\right),\quad
\sigma_2=\left(\begin{array}{cc}0&-i\\ i&0\end{array}\right),\quad
\sigma_3=\left(\begin{array}{cc}1&0\\ 0&-1\end{array}\right)
\label{e350}
\end{equation}
are Pauli matrices, we obtain the expression for the optimal unitary
operator:
\begin{eqnarray}
U_t=\left(\begin{array}{rr}\cos\left(\frac{\omega
t}{2\hslash}\right) & -{\rm i}
\sin\left(\frac{\omega t}{2\hslash} \right) \\ \\
-{\rm i}\sin\left( \frac{\omega t}{2\hslash}\right) &
\cos\left(\frac{\omega t} {2\hslash}\right)\end{array}\right).
\label{e24}
\end{eqnarray}
It follows at once that the optimal unitary orbit
$|\psi_t\rangle=U_t |\psi_I \rangle$ is given by
\begin{eqnarray}
|\psi_t\rangle=\left({\cos\left(\frac{\omega
t}{2\hslash}\right)\atop-{\rm i} \sin\left(\frac{\omega
t}{2\hslash}\right)}\right). \label{e25}
\end{eqnarray}
We find that the first time at which $|\psi_t\rangle$ reaches
$|\psi_F\rangle$ is given by the condition $\omega
t/2\hslash=\pi/2$, that is, when $t=\tau_{\rm P}$, where $\tau_{\rm
P}$ is the passage time given in (\ref{e15}).

We have seen how the simplest form of a quantum brachistochrone
problem can be solved in Hermitian quantum mechanics by considering
a two-dimensional Hilbert subspace combined with elementary
geometric constructions on it. In a more general situation the
unitary motion may be constrained further so that the optimal
Hamiltonian (\ref{e19}) may not be implementable. For example, the
constraint may enforce the path of the unitary evolution to lie in a
three- rather than in a two-dimensional subspace. To determine what
happens let us work out the passage time for this example.

Since in this case the minimal solution $H$ to the brachistochrone
problem is a three-dimensional matrix, we can express the initial
state $|\psi_I\rangle$ in terms of the three eigenstates of $H$
according to
\begin{eqnarray}
|\psi_I\rangle=\cos\alpha|E_i\rangle+\sin\alpha\cos\beta{\rm
e}^{{\rm i}\phi}| E_j\rangle+\sin\alpha\sin\beta{\rm e}^{{\rm
i}\varphi}|E_k\rangle, \label{e26}
\end{eqnarray}
where $\alpha,\,\beta$ are angular coordinates, $\phi,\,\varphi$ are
phase variables, and we assume that $E_i<E_j<E_k$. If a unitary
operator $U_T$ transforms this state into an orthogonal state, then
the condition
\begin{eqnarray}
\cos^2\alpha+\sin^2\alpha\cos^2\beta{\rm e}^{-{\rm
i}\omega_{ji}T/\hslash}+ \sin^2\alpha\sin^2\beta{\rm e}^{-{\rm
i}\omega_{ki}T/\hslash}=0 \label{e27}
\end{eqnarray}
must be satisfied, where $\omega_{ji}=E_j-E_i$ and
$\omega_{ki}=E_k-E_i$. To render the analysis more tractable, we
simplify this constraint by assuming that $\alpha=\beta=\pi/ 4$.
Then (\ref{e27}) implies that a necessary condition for the state
$|\psi_I\rangle$ to evolve into an orthogonal state is given by the
relation
\begin{eqnarray}
\frac{\omega_{ki}}{\omega_{ji}}=\frac{2m-1}{2n-1},
\label{e28}
\end{eqnarray}
where $m,n$ are natural numbers such that $m\neq n$.

Thus, the solution to the brachistochrone problem must be such that
the eigenvalues of $H$ satisfy condition (\ref{e28}) as well as the
constraint $E_{ \rm max}-E_{\rm min}\leq\omega$. Assuming that these
constraints are indeed satisfied, the initial state evolves into an
orthogonal state $|\psi_F\rangle$. The first time that
$|\psi_I\rangle$ evolves into a state orthogonal to $|\psi_I
\rangle$, in particular, is given by
\begin{eqnarray}
T=\frac{\pi\hslash}{\omega_{ji}}=\frac{3\pi\hslash}{\omega_{ki}}.
\label{e29}
\end{eqnarray}
However, since in this case $U_t|\psi_I\rangle$ does not describe a
geodesic path, $T$ will be larger than Fleming's passage time
$\tau_{\rm P}$ given in (\ref{e15}). Indeed, without loss of
generality we may set $E_i=0$. Then, it is straightforward to verify
that $T=\sqrt{6}\tau_{\rm P}$. This follows from the fact that under
the constraint $\omega_{ki}=3\omega_{ji}$ that comes from
(\ref{e29}), the squared energy dispersion in the state (\ref{e26})
with $\alpha =\beta=\pi/4$ is given by $\Delta
H^2=\frac{3}{2}\omega_{ji}^2$.

\section{Non-Hermitian Quantum Brachistochrone}
\label{s5}

We have seen how the solution to the simple brachistochrone problem
can be obtained in the Hermitian quantum theory. What happens if we
extend the quantum theory into the complex domain by looking at a
$PT$-symmetric theory? We saw earlier that in classical mechanics if
we were to allow for a complex path interpolating a pair of real
points of the coordinate space, then it is possible (at least
mathematically) to transport a particle across a large distance in
virtually no time. It turns out that an analogous situation emerges
in the $P T$-symmetric theory. Here we present a simple algebraic
calculation of the optimal evolution time from an initial state to a
final state by using a simple $2\times2$ Hamiltonian. As we have
remarked above, the $2\times2$ model suffices to cover general cases
because in the case of our simple brachistochrone problem the
solution is found on the two-dimensional subspace of the Hilbert
space spanned by the initial state $|\psi_I\rangle$ and the final
state $|\psi_F \rangle$. In the case of a $PT$-symmetric Hamiltonian
the variational approach gives a more direct way to handle the
brachistochrone problem. Thus, we shall first briefly revisit the
Hermitian case but expressed in the variational formalism and then
we will compare the result to its $PT$-symmetric counterpart.

\subsection{Hermitian $2\times2$ matrices}
\label{ss5.1}

We choose a basis so that the initial and final state vectors take
the form
\begin{equation}
|\psi_I\rangle=\left(\begin{array}{c}1\\0\end{array}\right)\quad{\rm
and}\quad |\psi_F\rangle=\left(\begin{array}{c}a\\
b\end{array}\right), \label{e30}
\end{equation}
where the condition that $|\psi_F\rangle$ be normalised is
$|a|^2+|b|^2=1$. The most general $2\times2$ Hermitian Hamiltonian
is
\begin{eqnarray}
H=\left(\begin{array}{cc} s & r\,\re^{-{\rm i}\theta}\cr
r\,\re^{{\rm i}\theta}&
u\end{array}\right)\qquad(r,~s,~u,~\theta~{\rm real}). \label{e31}
\end{eqnarray}
For this Hamiltonian the eigenvalue constraint (\ref{e4}) takes the form
\begin{eqnarray}
\label{e32}
\omega^2=(s-u)^2+4r^2.
\end{eqnarray}

To find the optimal Hamiltonian satisfying this constraint, we
rewrite $H$ as a linear combination of Pauli matrices:
\begin{equation}
H=\half(s+u){\mathds 1}+\half\omega\sigma\!\cdot\!{\bf n},
\label{e33}
\end{equation}
where
\begin{equation}
{\bf n}=\frac{1}{\omega}(2r\cos\theta,2r\sin\theta,s-u)
\label{e34}
\end{equation}
is a unit vector. Then by use of the identity (\ref{e23}) the
evolution equation $|\psi_F\rangle=\re^{-{\rm
i}H\tau/\hslash}|\psi_I\rangle$ can be expressed in the form
\begin{eqnarray}
\left(\begin{array}{c}a\\b\end{array}\right)=\re^{-\frac{1}{2}{\rm
i}(s+u)t/ \hslash}\left(\begin{array}{c}\cos\frac{\omega
t}{2\hslash}-{\rm i}\frac{s-u}{ \omega}\sin\frac{\omega
t}{2\hslash}\\ {}\\-{\rm i}\frac{2r}{\omega}\re^{{\rm i}
\theta}\sin\frac{\omega t}{2\hslash}\end{array}\right). \label{e37}
\end{eqnarray}
The second component of this equation gives
$|b|=\frac{2r}{\omega}\sin\frac{ \omega t}{2\hslash}$, which allows
us to find the required time of evolution:
\begin{equation}
t=\frac{2\hslash}{\omega}\arcsin\frac{\omega|b|}{2r}.
\label{e38}
\end{equation}

We must now minimise the time $t$ over all $r>0$ while maintaining
the energy constraint in (\ref{e32}). This constraint tells us that
the maximum value of $r$ is $\half\omega$. At this value we have
$s=u$. Because $H$ can be made trace free, we can set $s=u=0$. The
variable $\theta$ in (\ref{e33}) does not affect the eigenvalues, so
we may set $\theta=\pi$. Then we recover the optimal Hamiltonian
obtained in (\ref{e22}). As regards the minimum evolution time
$\tau$ we have
\begin{eqnarray}
\tau=\frac{2\hslash\arcsin|b|}{\omega}.
\label{e39}
\end{eqnarray}
In the special case for which $a=0$ and $b=1$ so that
$|\psi_I\rangle$ and $|\psi_F\rangle$ are orthogonal, we recover the
passage time $\tau=\tau_{\rm P}= \pi \hslash/\omega$, the smallest
time required for a spin flip.

Although the form of the result in (\ref{e39}) resembles the
statement of the uncertainty principle, it is merely the statement
indicated above that {\it rate $\times$time$=$distance}; the maximum
speed of evolution is given by $\Delta H$, and the distance between
$|\psi_I\rangle$ and $|\psi_F\rangle$, assuming they are normalised,
is given by $2\arccos(|\langle\psi_F|\psi_I\rangle|)$. Since
$|\langle\psi_F|\psi_I\rangle|=|a|$ and $|a|=\sqrt{1-|b|^2}$, we
obtain (\ref{e39}) from the relation
\begin{eqnarray}
{\rm arccos}\sqrt{1-|b|^2}={\rm arcsin}|b|.
\label{e39.7}
\end{eqnarray}

\subsection{Non-Hermitian $2\times 2$ matrices}
\label{ss5.2}

We now show by direct calculation that for a $PT$-symmetric
Hamiltonian, $\tau$ can be arbitrarily small. This is because a
$PT$-symmetric Hamiltonian whose eigenvalues are all real is
equivalent to a Hermitian Hamiltonian via $\tilde H=
\re^{-Q/2}H\re^{Q/2}$, where $Q$ is Dirac Hermitian. The states in a
$P T$-symmetric theory are mapped by $\re^{-Q/2}$ to the
corresponding states in the Dirac Hermitian theory. But, the overlap
distance between two states does not remain constant under a
similarity transformation. We can exploit this property of the
similarity transformation to overcome the Hermitian lower limit on
the time $\tau$. (The detailed calculation is explained in
Ref.~\cite{R38}.)

We consider the general class of $PT$-symmetric $2\times2$
Hamiltonians having the form
\begin{eqnarray}
H=\left(\begin{array}{cc}r\,\re^{{\rm i}\theta}&s\cr s &
r\,\re^{-{\rm i}\theta}
\end{array}\right)\qquad(r,~s,~\theta~{\rm real}).
\label{e40}
\end{eqnarray}
The time reversal operator $T$ performs complex conjugation and the
parity operator in this case is given by
\begin{eqnarray}
P=\left({0~~1}\atop{1~~0}\right).
\label{e40.2}
\end{eqnarray}
The two eigenvalues
\begin{equation}
E_\pm=r\cos\theta\pm\sqrt{s^2-r^2\sin^2\theta}
\label{e41}
\end{equation}
are real if $s^2>r^2\sin^2\theta$. This inequality defines the
region of unbroken $PT$ symmetry. The unnormalised eigenstates of
$H$ are
\begin{eqnarray}
|E_+\rangle=\left(\begin{array}{c}\re^{{\rm i}\alpha/2} \cr
\re^{-{\rm i}\alpha /2}\end{array}\right)\quad{\rm and}\quad
|E_-\rangle=\left(\begin{array}{c}{\rm i}\re^{-{\rm
i}\alpha/2}\cr-{\rm i}\re^{{\rm i}\alpha/2}\end{array}\right),
\label{e42}
\end{eqnarray}
where $\alpha$ is given by $\sin\alpha=(r/s)\sin\theta$. Note that
the condition of unbroken $PT$ symmetry of $H$ in (\ref{e40})
implies that $\alpha$ is real. The $C$ operator required for
defining the Hilbert space inner product is
\begin{eqnarray}
C=\frac{1}{\cos\alpha}\left(\begin{array}{cc}{\rm i}\sin\alpha & 1
\cr 1 & -{ \rm i}\sin\alpha\end{array}\right). \label{e43}
\end{eqnarray}
It is easy to verify that the $CPT$ norms of both eigenstates have the
value $\sqrt{2\cos\alpha}$.

To calculate $\tau$ we express the Hamiltonian $H$ in (\ref{e40}) as
\begin{equation}
H=(r\cos\theta){\mathds 1}+\half\omega\sigma\!\cdot\!{\bf n},
\label{e44}
\end{equation}
where
\begin{equation}
{\bf n}=\frac{2}{\omega}(s,0,{\rm i}r\sin\theta)
\label{e45}
\end{equation}
is a unit vector. The energy constraint requires that the squared
difference between energy eigenvalues is
\begin{eqnarray}
\omega^2=4s^2-4r^2\sin^2\theta.
\label{e46}
\end{eqnarray}
The positivity of $\omega^2$ is ensured by the condition of unbroken
$PT$ symmetry. Notice that (\ref{e46}) differs from (\ref{e32}) by a
sign. We can think of (\ref{e46}) as being \textit{hyperbolic} in
character, while (\ref{e32}) is \textit{elliptic} in character. The
technical advantage of the constraint in (\ref{e46}) is that because
of the minus sign both terms on the right side can become large
without violating the condition that $\omega$ be fixed. We will see
that it is this fact that allows the non-Hermitian Hamiltonian $H$
in (\ref{e40}) to achieve faster-than-Hermitian time evolution.

To determine $\tau$ we write down the $PT$-symmetric time-evolution
equation in vector form:
\begin{eqnarray}
\re^{-{\rm i}Ht/\hslash}\left(\begin{array}{c}
1\\0\end{array}\right)=\frac{ \re^{-{\rm
i}tr\cos\theta/\hslash}}{\cos\alpha}\left(\begin{array}{c}\cos(\frac{
\omega t}{2\hslash} -\alpha)\\ \\ -{\rm i}\sin\left(\frac{\omega
t}{2\hslash} \right)\end{array}\right). \label{e47}
\end{eqnarray}
In particular, consider the pair of vectors used in the Hermitian
spin-flip case as in (\ref{e20}). Observe that the evolution time
$t$ needed to reach $|\psi_F \rangle=\left({0\atop1}\right)$ from
$|\psi_I\rangle=\left({1\atop0}\right)$ is given by
\begin{eqnarray}
t=\frac{(2\alpha-\pi)\hslash}{\omega}.
\label{e47.3}
\end{eqnarray}
Optimising this result over allowable values for $\alpha$ as
$\alpha$ approaches $\half\pi$, the optimal time $\tau$ tends to
zero, a dramatic change from the Hermitian result in (\ref{e15})!
Note, however, that the two vectors in (\ref{e20}) are not
orthogonal with respect to the $CPT$ inner product. This is the
reason that the Fleming bound in (\ref{e15}) is not violated.

\section{Extension of Non-Hermitian Hamiltonians to
Higher-Dimensional Hermitian Hamiltonians} \label{s6}

We have seen how a quantum state can be transported unitarily into
another state in arbitrary short time by using a bounded Hamiltonian
if we allow for a complex path interpolating them in the space of
unitary motions. Can such an operation be implemented in practice?
If the answer is affirmative, then the implication is immense in
quantum information, computation, cryptography, and other related
fields. For example, if a quantum computer were to exist, then
solutions to difficult optimisation problems can in principle be
found in arbitrary short time, and this in turn would have important
implications in society as a whole.

A \textit{gedanken} experiment was proposed in Ref.~\cite{R38} to
realise this effect in a laboratory. The setup is as follows: we use
a Stern-Gerlach filter to create a beam of spin-up electrons. The
beam then passes through a \textit{black box} containing a device
governed by a $PT$-symmetric Hamiltonian that flips the spins
unitarily in a very short time. The outgoing beam then enters a
second Stern-Gerlach device that verifies that the electrons are now
in spin-down states. In effect, the black box device is
\textit{applying a complex magnetic field ${\vec B}$}:
\begin{eqnarray}
{\vec B}=(s,0,{\rm i}r\sin\theta).
\label{e48}
\end{eqnarray}
If the field strength has sufficiently large amplitude, then spins
can be flipped in virtually no time because the complex path joining
these two states is arbitrarily short without violating the energy
constraint in (\ref{e46}). We emphasise that the field strength can
be made large without violating the energy constraint (\ref{e4}) is
a consequence of the hyperbolic representation in (\ref{e46}).

The arbitrarily short alternative complex pathway from an up state
to a down state, as illustrated by this thought experiment, is
reminiscent of the short alternative distance between two widely
separated space-time points as measured through a wormhole in
general relativity. This comparison is of course controversial, and
it has subsequently motivated much research and it has generated a
lively debate in the literature
\cite{R41,R55,R50,R53,R57,R51,R52,R56,R54}. We emphasise that the
entire package of flipping the spin is not realised by a unitary
operation. This follows from the fact that $PT$-symmetric quantum
theory is unitary, and as such it respects the Fleming bound
(\ref{e15}) applicable to all unitary theories \cite{R38}. The point
is that the `black box' scheme described above actually consists of
three regimes: (i) the preparation of a spin-up state in the
Hermitian setup, (ii) the fast unitary motion to flip the spin using
a $PT$-symmetric Hamiltonian, and (iii) the recovery of a spin-down
state in the Hermitian setup. Thus, the operation is locally
unitary, but the switching between Hermitian and $P T$-symmetric
description is not unitary. This three-step process of switching
Hamiltonians is analogous to the classical procedure described in
Sec.~\ref{s3} for obtaining faster-than-real time evolution. Recall
that in the classical case we were able to transport a finite-energy
particle across a large distance in a short time by switching the
potential through which it was travelling. Note that in the
classical case there is no question of violating unitarity because
the particle does not get lost.

The question of unitarity and faster-than-Hermitian time evolution
has been reexamined in more detail in Ref.~\cite{R53,R56} by means
of a geometric approach and also in Ref.~\cite{R41}, where a more
general class of non-Hermitian Hamiltonians that are not necessarily
$PT$ symmetric are considered. In particular, Mostafazadeh has
emphasised the role of quantum observables in such an experiment;
the spin operator in Hermitian quantum mechanics cannot be
interpreted as a spin operator in the $PT$-symmetric counterpart,
thus leading to ambiguities regarding the physical interpretation of
the \textit{gedanken} experiment described above.

An intriguing alternative proposal for an implementation of the fast
spin flip has been made more recently by G\"unther and Samsonov
\cite{R57}. The idea is to embed the problem into a Hermitian setup
represented by a higher-dimensional Hilbert space. Specifically,
take our $PT$-symmetric Hamiltonian $H$ in (\ref{e40}). The
eigenstates of $H$ are not orthogonal with respect to the Hermitian
inner product. Since $H$ is not Hermitian, its Hermitian conjugate
defines a new matrix $H^\dagger$. The eigenstates of $H^\dagger$
thus also define a pair of nonorthogonal states in the Hermitian
theory. When these four states are suitably normalised, they can be
used to form an over-complete basis set in the Hermitian
two-dimensional Hilbert space. Such an over-complete set of basis is
also known as a positive operator-valued measure (POVM), commonly
used in the analysis of quantum information theory. A key idea is
that such a basis can be embedded in a higher-dimensional Hilbert
space to form an orthogonal basis by using the Naimark dilation
\cite{R49}. A Hermitian Hamiltonian can then be constructed---in
this case a $4\times4$ matrix---such that its eigenstates are
precisely the four states thus obtained. Using this Hamiltonian it
is possible to construct a standard unitary motion in such a way
that the induced motion obtained by the projection onto the
two-dimensional subspace is characterised by the $PT$-symmetric
motion (\ref{e47}).

In this way, G\"unther and Samsonov were able to show that the fast
spin flip can in principle be realised in the standard Hermitian
quantum mechanics by a combination of a unitary motion and a
projection in a larger-dimensional Hilbert space. In practical terms
this means that one should couple the spin to an auxiliary particle
(this can be done either by a projection or by a unitary operation),
apply a unitary evolution in the larger Hilbert space of the
combined system, and finally project out the auxiliary particle to
recover the spin in the transported state. The net effect of such an
operation can then be characterised by (\ref{e47}). The Fleming
bound is apparently violated due to the general fact that when a
unitary motion is projected to a subspace of a Hilbert space, the
resulting dynamics need not respect laws of unitarity. It would be
of considerable interest to find out whether the G\"unther and
Samsonov scheme can actually be implemented in a laboratory, and if
not, what might be the difficulty preventing the violation of the
Fleming bound.

\begin{acknowledgements}
We have benefited greatly from many discussions with
Drs.~U.~G\"unther and B.~Samsonov. We thank Dr.~D.~W.~Hook for his
assistance in preparing the figures used in this chapter. CMB is
supported by a grant from the U.S.~Department of Energy.
\end{acknowledgements}



\begin{thebibliography}{99.}

\bibitem{R47} Anandan, J.~and Aharonov, Y.: Geometry of quantum Eeolution.
Phys.~Rev.~Lett.~{\bf 65}, 1697-1700 (1990)

\bibitem{R41} Assis, P.~and Fring, A.: The quantum brachistochrone problem
for non-Hermitian Hamiltonians. J.~Phys.~A: Math.~Theor.~{\bf 41}, 244002
(12 pages) (2008)

\bibitem{R26} Barton, G.: \textit{Introduction to Advanced Field Theory}
(John Wiley \& Sons, New York, 1963), chap.~12

\bibitem{R2} Bender, C.~M.: Introduction to PT-symmetric quantum theory.
Contemp.~Phys.~{\bf 46}, 277-292 (2005)

\bibitem{R3} Bender, C.~M.: Making sense of non-Hermitian Hamiltonians.
Rep.~Prog.~Phys.~{\bf 70}, 947-1018 (2007)

\bigskip

\bibitem{R1} Bender, C.~M. and Boettcher, S.: Real spectra in Nnn-Hermitian
Hamiltonians having $PT$ symmetry. Phys.~Rev.~Lett.~{\bf 80}, 5243-5246 (1998)

\bibitem{R37} Bender, C.~M., Boettcher, S., and Meisinger, P.~N.:
$PT$-symmetric quantum mechanics. J.~Math.~Phys.~{\bf 40}, 2201-2209 (1999)

\bibitem{R10} Bender, C.~M., Brandt, S.~F., Chen, J.-H., and Wang, Q.: The $C$
operator in $PT$-symmetric quantum field theory transforms as a Lorentz scalar.
Phys.~Rev.~D {\bf 71}, 065010 (7 pages) (2005)

\bibitem{R25} Bender, C.~M., Brandt, S.~F., Chen, J.-H., and Wang, Q.: Ghost
busting: $PT$-Symmetric Interpretation of the Lee Model. Phys.~Rev.~D {\bf 71},
025014 (11 pages) (2005)

\bibitem{R36} Bender, C.~M., Brody, D.~C., Chen, J.-H., Jones, H.~F., Milton,
K.~A., and Ogilvie, M.~C.: Equivalence of a complex $PT$-symmetric quartic
Hamiltonian and a Hermitian quartic Hamiltonian with an anomaly. Phys.~Rev.~D
{\bf 74}, 025016 (10 pages) (2006)

\bigskip

\bibitem{R7} Bender, C.~M., Brody, D.~C., and Jones, H.~F.: Complex extension of
quantum mechanics. Phys.~Rev.~Lett.~{\bf 89}, 270401 (4 pages) (2002)

\bibitem{R8} Bender, C.~M., Brody, D.~C., and Jones, H.~F.: Must a Hamiltonian
be Hermitian? Am.~J.~Phys.~{\bf 71}, 1095-1102 (2003)

\bibitem{R9} Bender, C.~M., Brody, D.~C., and Jones, H.~F.: Scalar quantum field
theory with complex cubic interaction. Phys.~Rev.~Lett.~{\bf 93}, 251601 (4
pages) (2004)

\bibitem{R38} Bender, C.~M., Brody, D.~C., Jones, H.~F., and Meister, B.~K.:
Faster than Hermitian quantum mechanics. Phys.~Rev.~Lett.~{\bf 98}, 040403
(4 pages) (2007)

\bibitem{R55} Bender, C.~M., Brody, D.~C., Jones, H.~F., and Meister, B.~K.:
Comment on the quantum brachistochrone problem. arXiv:0804.3487 [quant-ph]

\bigskip

\bibitem{R27} Bender, C.~M.~and Mannheim, P.~D.: No-ghost theorem for the
fourth-order derivative Pais-Uhlenbeck oscillator model.
Phys.~Rev.~Lett.~{\bf 100}, 110402 (4 pages) (2008)

\bibitem{R28} Bender, C.~M.~and Mannheim, P.~D.: Exactly solvable $PT$-symmetric
Hamiltonian having no Hermitian counterpart. Phys.~Rev.~D {\bf 78}, 025022
(17 pages) (2008)

\bibitem{R29} Bender, C.~M.~and Mannheim, P.~D.: Giving up the Ghost.
J.~Phys.~A: Math.~Theor.~{\bf 41}, 304018 (7 pages) (2008)

\bibitem{R42} Brody, D.~C.: Elementary derivation of passage times.
J.~Phys.~A: Math.~Gen.~{\bf 36}, 5587-5593 (2003)

\bibitem{R40} Brody, D.~C.~and Hook, D.~W.: On optimum Hamiltonians for state
transformations. J.~Phys.~A: Math.~Gen.~{\bf 39}, L167-L170 (2006).
[Corrigendum. \textit{Ibid.}~{\bf 40}, 10949 (2007)]

\bigskip

\bibitem{R43} Brody, D.~C.~and Hughston, L.~P.: Geometric quantum mechanics.
J.~Geom.~Phys.~{\bf 38}, 19-53 (2001)

\bibitem{R17} Brower, R., Furman, M., and Moshe, M.: Critical exponents for the
Reggeon quantum spin model. Phys.~Lett.~B {\bf 76}, 213-219 (1978)

\bibitem{R34} Buslaev, V.~and Grecchi, V.: Equivalence of unstable anharmonic
oscillators and double wells. J.~Phys.~A: Math.~Gen.~{\bf 26}, 5541-5549 (1993)

\bibitem{R14} Cardy, J.~L.: Conformal invariance and the Yang-Lee edge
singularity in two dimensions. Phys.~Rev.~Lett.~{\bf 54}, 1354-1356 (1985)

\bibitem{R15} Cardy, J.~L.~and Mussardo, G.: S-matrix of the Yang-Lee edge
singularity in two dimensions. Phys.~Lett.~B {\bf 225}, 275-278 (1989)

\bigskip

\bibitem{R39} Carlini, A., Hosoya, A., Koike, T., and Okudaira, Y.: Time-optimal
quantum evolution. Phys.~Rev.~Lett.~{\bf 96}, 060503 (4 pages) (2006)

\bibitem{R5} Dorey, P., Dunning, C., and Tateo, R.: Supersymmetry and the
spontaneous breakdown of $PT$ symmetry. J.~Phys.~A: Math.~Gen.~{\bf 34},
L391-L400 (2001)

\bibitem{R6} Dorey, P., Dunning, C., and Tateo, R.: Spectral equivalences, Bethe
ansatz equations, and reality properties in $PT$-symmetric quantum mechanics.
J.~Phys.~A: Math.~Gen.~{\bf 34}, 5679-5704 (2001)

\bibitem{R4} Dorey, P., Dunning, C., and Tateo, R.: The ODE/IM correspondence.
J.~Phys.~A: Math.~Theor.~{\bf 40}, R205-R283 (2007)

\bibitem{R22} Faria, C.~F.~de M.~and Fring, A.: Non-Hermitian Hamiltonians
with real eigenvalues coupled to electric fields: from the time-independent to
the time dependent quantum mechanical formulation. Laser Phys.~{\bf 17}, 424-437
(2007)

\bigskip

\bibitem{R13} Fisher, M.~E.: Yang-Lee edge singularity and $\phi^3$
field theory. Phys.~Rev.~Lett.~{\bf 40}, 1610-1613 (1978)

\bibitem{R45} Fleming, G.~N.: A unitary bound on the evolution of nonstationary
states. Nuov.~Cim.~A {\bf 16}, 232 (1973)

\bibitem{R50} G\"unther, U., Rotter, I., and Samsonov, B.~F.: Projective Hilbert
space structures at exceptional points. J.~Phys.~A: Math.~Theor.~{\bf 40},
8815-8833 (2007)

\bibitem{R21} G\"unther, U., Samsonov, B.~F., and Stefani, F.: A globally
diagonalizable $\alpha^2$-dynamo operator, SUSY QM and the Dirac equation.
J.~Phys.~A: Math.~Theor.~{\bf 40}, F169-F176 (2007)

\bibitem{R53} G\"unther, U.~and Samsonov, B.~F.: The PT-symmetric
brachistochrone problem, Lorentz boosts and non-unitary operator equivalence
classes. arXiv: 0709.0483 [quant-ph] 

\bigskip

\bibitem{R57} G\"unther, U.~and Samsonov, B.~F.: The Naimark dilated $P
T$-symmetric brachistochrone. arXiv:0807.3643

\bibitem{R20} Guenther, U., Stefani, F., and Znojil, M.: MHD $\alpha^2$-dynamo,
Squire equation and $PT$-symmetric interpolation between square well and
harmonic oscillator. J.~Math.~Phys.~{\bf 46}, 063504 (22 pages) (2005)

\bibitem{R18} Harms, B., Jones, S., and Tan, C.-I: New structure in the energy
spectrum of Reggeon quantum mechanics with quartic couplings. Phys.~Lett.~B {\bf
91}, 291-295 (1980)

\bibitem{R19} Harms, B., Jones, S., and Tan, C.-I: Complex energy spectra in
Reggeon quantum mechanics with quartic interactions. Nucl.~Phys.~B {\bf 171},
392-412 (1980)

\bibitem{R49} Holevo, A.~S.: \textit{Probabilistic and Statistical Aspects of
Quantum Theory} (North-Holland, Amsterdam, 1982)

\bigskip

\bibitem{R12} Hollowood, T.: Solitons in affine Toda field theories.
Nucl.~Phys.~B {\bf 384}, 523-540 (1992)

\bibitem{R44} Hughston, L.~P.: Geometric aspects of quantum mechanics. In
\textit{Twistor Theory}, ed. by S.~Huggett (Marcel Dekker, New York, 1995)

\bibitem{R35} Jones, H.~F.~and Mateo, J.: Equivalent Hermitian Hamiltonian for
the non-Hermitian $-x^4$ potential. Phys.~Rev.~D {\bf 73}, 085002 (4 pages)
(2006)

\bibitem{R24} K\"all\'en, G.~and Pauli, W.: On the mathematical structure of
T.~D.~Lee's model of a renormalizable field theory. Dansk
Vid.~Selsk.~Mat.-Fys.~Medd.~{\bf 30}, No.~7, 23 pages (1955)

\bibitem{K2} Kibble, T.~W.~B.: Geometrization of quantum mechanics.
Commun.~Math.~Phys.~{\bf 65}, 189-201 (1979)

\bigskip

\bibitem{R48} Kobayashi,~S.~and Nomizu,~K.: \textit{Foundations of Differential
Geometry}, Vol. 2 (Wiley, New York, 1969)

\bibitem{R23} Lee, T.~D.: Some special examples in renormalizable field theory.
Phys.~Rev.~{\bf 95}, 1329-1334 (1954)

\bibitem{R31} Makris, K.~G., El-Ganainy, R., Christodoulides, D.~N., and
Musslimani, Z.~H.: Beam dynamics in PT symmetric optical lattices.
Phys.~Rev.~Lett.~{\bf 100}, 103904 (4 pages) (2008)

\bibitem{R51} Martin, D.: Is $PT$-symmetric quantum mechanics just quantum
mechanics in a non-orthogonal basis? arXiv: quant-ph/0701223v2

\bibitem{X9} Mostafazadeh,~A.: Pseudo-Hermiticity versus $PT$-symmetry III.
J.~Math.~Phys.~{\bf 43}, 3944-3951 (2002)

\bigskip

\bibitem{R33} Mostafazadeh, A.: Exact PT-symmetry is equivalent to Hermiticity.
J.~Phys.~A: Math.~Gen.~{\bf 36} (2003), 7081-7091.

\bibitem{R52} Mostafazadeh, A.: Quantum brachistochrone problem and the geometry
of the state space in pseudo-Hermitian quantum mechanics. Phys.~Rev.~Lett.~{\bf
99}, 130502 (4 pages) (2007)

\bibitem{R56} Mostafazadeh, A.: Physical meaning of Hermiticity and shortcomings
of the composite (Hermitian + non-Hermitian) quantum theory of G"unther and
Samsonov. arXiv:0709.1756 [quant-ph]

\bibitem{R30} Musslimani, Z.~H., Makris, K.~G., El-Ganainy, R., and
Christodoulides, D.~N.: Analytical solutions to a class of nonlinear
Schr\"odinger equations with $PT$-like potentials. J.~Phys.~A:
Math.~Theor.~{\bf 41}, 244019 (12 pages) (2008)

\bibitem{R32} Musslimani, Z.~H., Makris, K.~G., El-Ganainy, R., and
Christodoulides, D.~N.: Optical solitons in $PT$ periodic potentials.
Phys.~Rev.~Lett.~{\bf 100}, 030402 (4 pages) (2008)

\bigskip

\bibitem{R54} Rotter, I.: The brachistochrone problem in open quantum systems.
J.~Phys.~A: Math.~Theor.~{\bf 40} 14515-14526 (2007)

\bibitem{R58} Salam, A.: Review of \textit{On the mathematical structure of
T.~D.~Lee's model of a renormalizable field theory} by G.~K\"all\'en and
W.~Pauli, MathSciNet Mathematical Reviews on the Web MR0076639 (17,927d) (1956)

\bibitem{R46} Schulman, L.~S.: Jump Time and Passage Time. In \textit{Time
in Quantum Mechanics}, ed. by J.~G.~Muga, R.~Sala~Mayato, and I.~L.~Egusquiza
(Springer-Verlag, Berlin, 2002)

\bibitem{X7} Wigner,~E.~P.: Phenomenological distinction between unitarity and
antiunitarity symmetry operators. J.~Math.~Phys.~{\bf 1}, 414-416 (1960)

\bibitem{R11} Wu, T.~T.: Ground state of a Bose system of hard spheres.
Phys.~Rev.~{\bf 115}, 1390-1404 (1959)

\bigskip

\bibitem{R16} Zamolodchikov, A.~B.: Two-point correlation function in scaling
Lee-Yang model. Nucl.~Phys.~B {\bf 348}, 619-641 (1991)

\end{thebibliography}
\end{document}